\documentstyle[epsfig]{mn}

\begin{document}

\title{Evolutionary tracks of FRII sources through the P--D diagram}

\author[C. R. Kaiser, J. Dennett--Thorpe \& P. Alexander]{Christian
R. Kaiser\thanks{email: ckaiser@mrao.cam.ac.uk}, Jane Dennett--Thorpe and
Paul Alexander\\
MRAO, Cavendish Laboratory, Madingley Road, Cambridge CB3 0HE}

\maketitle

\begin{abstract}
This is the second in a series of papers presenting an analytical
model for the evolution of FRII radio sources. In this paper we
evaluate the expected radio emission from a radio source incorporating
energy loss processes for the relativistic electrons. By combining
these results with our earlier dynamical model we calculate
evolutionary tracks through the Power--Linear size diagram. These
tracks are in good agreement with the observed distribution of sources
in this diagram. The effects of different forms for the evolution of
the magnetic field in the cocoon, the redshift of the source, the
environment of the source and the defining parameters of the jet are
investigated. The evolutionary tracks are found to be insensitive to
the assumed form of the magnetic field evolution. Some evidence
against protons as a major constituent of the jet material is also
found. 
\end{abstract}


\section{Introduction}

This is the second in a series of papers presenting an analytical
model for the evolution of FRII radio sources; in the first paper
(Kaiser \& Alexander 1997, hereafter KA97\nocite{ka96b}) we
investigated the dynamical evolution of the source and here we extend
this analysis to consider the luminosity evolution of a radio source
and the tracks it follows through the Power--Linear size (P--D)
diagram. The P--D diagram was introduced by Shklovskii
(1963)\nocite{is63} as a powerful tool for investigating the temporal
evolution of FRI and FRII radio sources. By plotting the radio
luminosity at a specific frequency, $P_{\nu}$, as a function of the
linear size of a source, $D$, a diagram analogous to the
Hertzsprung--Russel diagram is obtained. The evolution of a radio
source through the P--D diagram is, however, not well understood. Some
constraints can be placed on these evolutionary tracks by the relative
densities of objects in regions of the P--D diagram; specifically,
sources with large linear sizes ($D>1$ Mpc) and high radio
luminosities ($P_{\nu} > 10^{26}$ W Hz$^{-1}$ sr$^{-1}$ at $\nu = 178$
MHz) seem to be rather rare suggesting that the luminosity of sources
should decrease quickly with linear size for sizes approaching 1
Mpc. Care must be taken in interpreting the P--D diagram in this way
since selection effects have a strong influence on the observed
high--redshift population (e.g. Masson 1980\nocite{cm80}, Macklin
1982\nocite{jm82}) and giant sources are likely to be difficult to
detect since the extended emission of the cocoon will be weak and the
hot spots may therefore appear unconnected (e.g. Muxlow \& Garrington
1991\nocite{mg91}). Searches specifically for giants have been
undertaken by Cotter {\em et al.} (1996)\nocite{crs96} and
Subrahmanyan {\em et al.} (1996)\nocite{ssh96}. Both groups identify a
number of radio sources in low frequency surveys as classical doubles
of extreme size, although almost all of these sources have
luminosities below $2\times 10^{26}$ W Hz$^{-1}$ sr$^{-1}$ suggesting
that there is a real deficit of powerful giant sources.

Baldwin (1982)\nocite{jb82} used the model for FRII sources of Scheuer
(1974)\nocite{ps74} and minimum energy arguments (e.g. Burbidge
1956\nocite{gb56}) to calculate evolutionary tracks through the P--D
diagram. In its original form his model predicts $P \propto D^{7/8}$
for a power law energy spectrum of the relativistic electrons, $N(E)\,
dE \propto E^{-p}\, dE$, with $p=2$ as exponent. This implies that all
sources would crowd together in exactly that area of the diagram which
is almost empty. Modifications of the model incorporating a density
gradient in the external medium surrounding the radio source ($\rho _x
\propto d^{-\beta}$, where $d$ is the distance from the center of the
distribution) predicted tracks with decreasing radio luminosity in
reasonable agreement with observations for $\beta=1.9$ and $D < 300$
kpc. To explain the rapid decline in luminosity for $D \approx 1$ Mpc
Baldwin suggested that the density of the atmosphere may decrease more
steeply with $\beta = 2.9$ for $D > 300$ kpc. However, sources in
environments with a density gradient steeper than $1/d^2$ are unlikely
to develop the typical FRII morphology (Falle 1991\nocite{sf91},
Kaiser \& Alexander 1997\nocite{ka96b}).

An alternative hypothesis to account for the decrease in luminosity at
large $D$ is that the central engine ceases to supply energy to the
jet after a certain period which is similar for all sources (for a
discussion see Daly 1994\nocite{rd94}). The cocoon will then dim
quickly via adiabatic expansion. Observations indicate that only of
order 4\% of all radio sources show no signs of ongoing AGN activity
Giovannini (1988)\nocite{gfgp88} in support of this hypothesis.

The synchrotron radiation of FRIIs demonstrates the presence of highly
relativistic electrons in the cocoon (e.g. Rees 1971\nocite{mr71}). To
achieve charge neutrality in the cocoon we need some species of
positively charged particles. If we assume that there is no
entrainment across the contact discontinuity which is separating the
cocoon from the IGM the jet must supply these additional
particles. The nature of these charge--balancing particles is subject
to debate (e.g. Leahy 1991\nocite{jl91}). If they are protons, a
significant fraction of the kinetic energy transported by the jet will
be stored by these in the cocoon but will not contribute to the
radiated power. Alternatively, if the particles providing charge
neutrality were positrons, they would behave almost exactly like
electrons in a completely tangled magnetic field and will contribute
to the observed synchrotron emission. The kinetic power of the jet
required to produce a cocoon of given radio luminosity would be
significantly reduced in the case of a pair plasma.

In this paper we extend the model described in KA97 for FRII sources
by the addition of energy loss processes of the relativistic electrons
in the cocoon to compute evolutionary tracks through the P--D
diagram. We show that changes in the external atmosphere or a
mechanism for switching off the central engine are not required to
reproduce the observed distribution of sources in the P--D diagram. We
also find some evidence against protons as a major constituent of the
jet material. In section \ref{THEO} we discuss synchrotron emission,
the energy loss processes of the relativistic electrons responsible
for this emission and a model for the cocoon based on
KA97\nocite{ka96b} which allows us to calculate its radio
emission. Section \ref{DISC} contains a discussion of the results
obtained with this model for the cocoon and examines changes to the
evolutionary tracks through the P--D diagram introduced by different
models for the evolution of the magnetic field in the cocoon, the
redshift of the source, its orientation with respect to the line of
sight and the addition of thermal, non--radiating particles to the
material in the cocoon.

\section{The radio emission of the cocoon\label{theo}}

In this section we firstly present a brief review of the synchrotron
emission mechanism and the energy losses affecting the relativistic
electrons responsible for this emission. We will assume these
electrons to be confined to a small volume element. We then go on to
develop a model for the emissivity of the whole of the cocoon by
summing up the contribution of each volume element. We neglect the
emission of the hot spots since we are confining our interest in this
paper to those radio sources in which the emission of the cocoon
dominates the total radio luminosity of the source.

\subsection{Synchrotron radiation and loss processes\label{loss}}

The specific volume emissivity due to synchrotron radiation,
$j_{\nu}$, of an ensemble of relativistic electrons in a magnetic
field of energy density $u_{\scriptscriptstyle B}$ averaged over all
electron pitch angles is (e.g. Shu 1991\nocite{fs91}):

\begin{equation} j_{\nu} = \int _1 ^{\infty} \frac{4}{3} \, \sigma _{\scriptscriptstyle T}
\, c \, u_{\scriptscriptstyle B} \, \beta _e ^2 \, \gamma ^2 \Phi (\nu,\gamma) \, n(\gamma) \;
d\gamma, \label{voem}\end{equation}

\noindent where $\sigma _{\scriptscriptstyle T}$ is the Thompson cross
section, $c$ the speed of light, $\beta _e$ the speed of the
relativistic electrons in units of $c$ ($\beta _{e} \approx 1$),
$\gamma$ the corresponding Lorentz factor and $n(\gamma) \, d\gamma$
the number density of electrons with Lorentz factors between $\gamma$
and $\gamma + d\gamma$. Taking the average over all possible electron
pitch angles is justified since we assume that the magnetic field is
completely tangled on scales much smaller than the cocoon itself. The
spectral emission of an electron of Lorentz factor $\gamma$ is given
by $\Phi(\nu,\gamma)$. In this paper we are primarily concerned with the
total source luminosity, rather then details of spectral shape and
therefore we use the standard approximation that electrons are
emitting only at their critical frequency $\nu = \gamma ^2 \nu
_{\scriptscriptstyle L}$, where $\nu_{\scriptscriptstyle L}$ is the
Larmor frequency; this allows us to set $\Phi(\nu,\gamma) = \delta
(\nu - \gamma ^2 \nu_{\scriptscriptstyle L})$. Taking the electron
ensemble to occupy a volume $V$ the emitted radio power per unit
frequency and solid angle is

\begin{equation} P_{\nu} = \frac{1}{6 \pi} \, \sigma _{\scriptscriptstyle
T} \, c \, u_{\scriptscriptstyle B} \, \frac{\gamma ^3}{\nu} \,
n(\gamma) \, V.\label{rapo}\end{equation}

If the relativistic electrons are initially accelerated at time $t_i$
(for example via a first order Fermi process at the jet
shock) we expect their energy distribution to be a power law function
of their initial Lorentz factor, $\gamma _i$, (e.g. Heavens \& Drury
1988)\nocite{hd88}

\begin{equation} n(\gamma _i, t_i) \, d\gamma _i = n_o \, \gamma _i
^{-p} \, d\gamma _i.\label{pola}\end{equation}

\noindent This spectrum will then evolve with time as the electron
population loses internal energy. The rate of change of the Lorentz
factor for the electron population is given by

\begin{equation} \frac{d\gamma}{dt} = -\frac{a_1}{3} \, \frac{\gamma}{t} -
\frac{4}{3} \, \frac{\sigma _{\scriptscriptstyle T}}{m_e c} \, \gamma ^2
\, \left( u_{\scriptscriptstyle B} + u _{\scriptscriptstyle C}
\right). \label{lora}\end{equation}

\noindent The first term on the right hand side represents adiabatic
expansion losses for the case where the volume $V$ expands as $V
\propto t^{a_1}$ (e.g. Longair 1981)\nocite{ml81}. The second term is
the combined loss rate due to synchrotron radiation and inverse
Compton scattering of the cosmic microwave background radiation
(CMBR); $m_e$ is the electron mass and $u_{\scriptscriptstyle C}$ is
the energy density of the CMBR. Integration of equation (\ref{lora})
gives

\begin{equation} \frac{t^{-\frac{a_1}{3}}}{\gamma} \, - \frac{t_i
^{-\frac{a_1}{3}}}{\gamma _i} = \frac{4}{3} \, \frac{\sigma
_{\scriptscriptstyle T}}{m_e c} \, \int _{t_i}^t \left(
u_{\scriptscriptstyle B} + u_{\scriptscriptstyle C} \right) \, \left(
t' \right) ^{-\frac{a_1}{3}} \, dt',\label{hawa}\end{equation}

\noindent where for an electron with a Lorentz factor of $\gamma$ at
time $t$, the Lorentz factor at time $t_i$, when it was initially
accelerated, is $\gamma _i$, and we may therefore consider $\gamma _i$
to be a function of $\gamma$ and $t$. As mentioned above we assume the
magnetic field to be completely tangled on all relevant scales,
allowing us to treat $u_{\scriptscriptstyle B}$ as a pressure with
adiabatic index, $\Gamma _{\scriptscriptstyle B}$, such that
$u_{\scriptscriptstyle B} \propto t^{-\Gamma _B a_1}$. The energy
density of the CMBR, $u_{\scriptscriptstyle C}$, is a function of
redshift ($u_{\scriptscriptstyle C} \propto (z+1)^4$), but since the
lifetime of radio sources does not exceed a few times 10$^8$ years
(Alexander \& Leahy 1987\nocite{al87}) we take $u_{\scriptscriptstyle
C}$ to be constant during the evolution of a given source. Performing
the integral in equation (\ref{hawa}):

\begin{equation} \frac{t^{-\frac{a_1}{3}}}{\gamma} -
\frac{t_i^{-\frac{a_1}{3}}}{\gamma _i} = a_2(t,t_i),
\label{loeq}\end{equation}

\noindent where 

\begin{eqnarray} \lefteqn{a_2(t,t_i) = \frac{4 \sigma _T}{3 m_e c}} \nonumber\\
& & \times \, \left[ \frac{u_{\scriptscriptstyle B}(t_i)}{a_3}\,
t_i^{a_1 \Gamma _{\scriptscriptstyle B}} \, \left( t^{a_3} - t_i^{a_3}
\right) + \frac{u_{\scriptscriptstyle C}}{a_4} \, \left( t^{a_4} -
t_i^{a_4}\right) \right],
\end{eqnarray}

\noindent with $a_3 = 1 - a_1 ( \Gamma _{\scriptscriptstyle
B} + 1/3)$ and $a_4 = 1 - a_1 / 3$.  If $u_e(t_i)$ is the energy
density of the relativistic electrons at time $t_i$, we find for
$n_o(t_i)$ from equation (\ref{pola})

\begin{eqnarray} \lefteqn{n_o =  \frac{u_e(t_i)}{m_e c^2} \left( \int _{\gamma _{i,min}} ^{\gamma
_{i,max}} \, (\gamma _i -1) \, \gamma _i^{-p} \; d\gamma _i\right)
^{-1}} \nonumber\\ & &= \frac{u_e(t_i)}{m_e c^2} \, \left(
\frac{\gamma _{i,min}^{2-p} - \gamma _{i,max}^{2-p}}{p-2} -
\frac{\gamma _{i,min}^{1-p} - \gamma _{i,max}^{1-p}}{p-1} \right)
^{-1}, \label{rede}\end{eqnarray}

\noindent where $\gamma _{i,min}$ and $\gamma _{i,max}$ are the low
and high energy cut--off of the initial energy distribution. The
electrons are uniformly distributed over the volume $V$. Because of
the expansion of $V$ we have to set $t^{a_1} n(\gamma,t) \, d\gamma =
t_i^{a_1} n(\gamma_i,t_i) \, d\gamma_i$. Thus the energy distribution of
the relativistic electrons at time $t$ can be shown to be

\begin{equation} n(\gamma , t) \, d\gamma = n_o \, \frac{\gamma _i 
^{2-p}}{\gamma ^2} \left( \frac{t}{t_i} \right) ^{-\frac{4a_1}{3}}
d\gamma,  \label{npola}\end{equation}

\noindent where from equation (\ref{loeq})

\begin{equation} \gamma _i  = \frac{\gamma \, t_i
^{-\frac{a_1}{3}}}{t^{-\frac{a_1}{3}} - a_2(t,t_i) \,
\gamma}\label{oga}\end{equation}

\noindent and $n_o$ is given by equation (\ref{rede}).

\subsection{A model for the cocoon}

The overall dynamics of the cocoon, and specifically the evolution of
the cocoon pressure, $p_c$, were considered in detail in KA97; in this
section we review the essentials of this model necessary to calculate
the luminosity of the source and to determine also how an element of
radio emitting plasma evolves within the cocoon.

For the calculation of the radio emission of the cocoon we divide its
contents into three separate `fluids' with individual energy
densities. The first fluid consists of the electrons (and possibly
positrons); these are described by the energy spectrum given in
equation (\ref{pola}) and contribute an energy density $u_e$ with
adiabatic index $\Gamma _e$. For a completely tangled field we have a
magnetic `fluid' with energy density $u_{\scriptscriptstyle B}$ and
adiabatic index, $\Gamma _{\scriptscriptstyle B}=4/3$. Finally, we allow
for the possibility that protons and/or electrons with a thermal
spectrum are present in the cocoon with an energy density
$u_{\scriptscriptstyle T}$ and adiabatic index $\Gamma
_{\scriptscriptstyle T}$. The overall dynamics are governed by the
pressure in the cocoon, $p_c=(\Gamma _c -1)(u_e+u_{\scriptscriptstyle
B}+u_{\scriptscriptstyle T})$; where the adiabatic index of the cocoon
as a whole, $\Gamma _c$, depends on the relative pressures of each
component.

To allow for variations of the energy densities within the cocoon we
split the cocoon into small volume elements $\delta V$. The fluid
present in the volume $\delta V$ is injected into the cocoon at a time
$t_i$ over a short time interval $\delta t_i$, during which the
internal energy of $\delta V$, $\delta U$, changes according to

\begin{equation} d(\delta U) = Q_o \, d(\delta t_i) - p_c(t_i) \, d (\delta
V),\label{expan}\end{equation}

\noindent where $Q_o$ is the power of the jet. During the interval
$\delta t_i$ the volume elements expand adiabatically which changes
their pressure from the hot spot pressure $p_h(t_i)$ to that of the
cocoon $p_c(t_i+\delta t_i) \approx p_c (t_i)$. With this assumption
integration of equation (\ref{expan}) gives

\begin{eqnarray} \delta U & = & \frac{p_c(t_i) \, \delta V (t_i)}{\Gamma _c
-1} \nonumber \\ & = & Q_o \, \delta t_i - \frac{p_c (t_i) \, \delta V
(t_i)}{\Gamma _c -1} \, \left[ \left( \frac{p_h (t_i)}{p_c (t_i)}
\right) ^{\frac{\Gamma _c -1}{\Gamma _c}} -1 \right].
\end{eqnarray}

\noindent From KA97 we take $p_h / p_c \approx 4 R^2$ which then yields

\begin{equation} \delta V (t_i) = \frac{(\Gamma _c -1) \, Q_o}{p_c(t_i)}
\, \left( 4 R^2 \right) ^{\frac{1-\Gamma _c}{\Gamma _c}} \, \delta t_i.
\end{equation}

\noindent If the expansion of $\delta V$ after the interval $\delta
t_i$ is also adiabatic we finally find

\begin{equation} \delta V (t)  =  \frac{\left( \Gamma
_c - 1 \right) \, Q_o}{p_c(t_i)} \, \left( 4 R^2 \right)
^{\frac{1-\Gamma _c}{\Gamma _c}} \, \left( \frac{t}{t_i} \right)
^{a_1} \, \delta t_i,\label{voeq}
\end{equation}

\noindent where $a_1 = (4+\beta) / (\Gamma _c (5-\beta))$. Here we
have used the result from KA97 that $p_c \propto
t^{(-4-\beta)/(5-\beta)}$ with $\beta$ being the exponent in the
density distribution of the external atmosphere.

We define $k'$ as the ratio of the energy densities of the thermal
particles, $u_{\scriptscriptstyle T}$, to that of the electrons,
$u_e$, when they are injected into the cocoon at $t_i$. Note that this
differs from the usual way of defining $k$ as the ratio of the energy
densities of non--radiating particles and relativistic electrons. In
our case the non--relativistic electrons of the power law energy
distribution (equation \ref{npola}) are already included in $u_e$. We
also introduce $r$ as the ratio of the energy density of the magnetic
field, $u_{\scriptscriptstyle B}$, to the sum of the energy densities
of the particles, $u_e + u_{\scriptscriptstyle T}$. With these
definitions:

\begin{eqnarray} u_e (t_i)& = &
\frac{p_c(t_i)}{(\Gamma _c -1) (k'+1) (r+1)}, \nonumber\\
u_{\scriptscriptstyle B} (t_i)& = & \frac{r \, p_c(t_i)}{(\Gamma _c
-1) (r+1)}.
\end{eqnarray}

By identifying $V$ in equation (\ref{rapo}) with $\delta V$ we can
calculate the radio emission of this volume element, and obtain the
total emission of the cocoon by summing the contributions of all such
elements within the cocoon; from equation (\ref{voeq}) it is clear
that this sum reduces to an integration over the injection time,
$t_i$:

\begin{eqnarray} \lefteqn{P_{\nu} = \int _0^t \frac{\sigma _{\scriptscriptstyle
T} \, c \, r}{6 \pi \, \nu \, (r+1)} \, Q_o \, n_o \, \left( 4 R^2
 \right) ^{\frac{1- \Gamma _c}{\Gamma _c}}}
 \nonumber\\ & \times & \frac{\gamma ^{3-p} \,
 t_i^{\frac{a_1}{3}(p-2)}}{\left( t^{- \frac{a_1}{3}} - a_2 (t,t_i) \,
 \gamma \right) ^{2-p}} \, \left( \frac{t}{t_i} \right) ^{-a_1 \left(
 \frac{1}{3} + \Gamma _{\scriptscriptstyle B} \right)} dt_i.
\label{beau}
\end{eqnarray}

For the older parts of the cocoon the energy losses of the
relativistic electrons can be so severe that $\gamma _i \rightarrow
\infty$ in equation (\ref{oga}), i.e. electrons with the correct
Lorentz factor, $\gamma$, to produce radiation with frequency $\nu$
now, at time $t$, would have had $\gamma _i \rightarrow \infty$ at
injection time, $t_i$. Volume elements for which this is the case no
longer contribute to the radio emission, hence we define a minimum
injection time, $t_{min}$, at which the cocoon material is still
radiating at frequency $\nu$. The integration limits in equation
(\ref{beau}) are then from $t_{min}$ to $t$.

The integral in equation (\ref{beau}) is not analytically soluble for
arbitrary $p$, and we therefore use a numerical approach, the results
of which we present in Section \ref{DISC}.

\section{Evolutionary tracks\label{disc}}

To perform the calculations discussed in section \ref{THEO} an
explicit form for the electron distribution function at the time the
electrons are injected into the cocoon is required. It is shown in
KA97 that the material in the jet is `cold' implying an adiabatic
index of 5/3. Nevertheless, in the rest frame of the jet shock the jet
material is moving at relativistic speeds (0.77c $\rightarrow$ 0.87c
for the parameters used in KA97), which implies an exponent for the
energy distribution of $p = 2.14$ (Heavens \& Drury
1988\nocite{hd88}). This value is derived using a test particle
approach which may not be appropriate for jet shocks in extragalactic
sources. However, the spectral indices of observed radio sources infer
$2 \le p \le 3$ (e.g. Alexander \& Leahy 1987\nocite{al87}) and we
therefore use $p=2.14$ for the tracks calculated in this
section. Since $p > 2$, all moments of the electron energy
distribution up to the second moment (related to the synchrotron loss
rate) are insensitive to the choice of upper limit for the energy
distribution, $\gamma _{i,max}$, and specifically converge even as
$\gamma _{i,max} \rightarrow \infty$; we therefore take this limiting
value. The main form in which energy is transported along the jet is
as bulk kinetic energy; the energy transported per particle in the jet
is therefore approximately constant. The acceleration process at the
jet shock transfers this energy to the fraction of the electron
population constituting the power law energy distribution; a
substantial proportion of the material in the cocoon must therefore be
cold, contributing to the density but not the energy density. So as to
account for this population we assume the power law distribution
extends to $\gamma _{i,min} =1$; however we shall for the moment
neglect any thermal plasma setting $k'=0$.

From KA97 we have expressions for the cocoon pressure $p_c$ (assuming
cylindrical symmetry of the cocoon) and for the linear size as a
function of time ($D \propto t^{3/(5-\beta)}$). For the initial ratio
of the energy densities of the magnetic field and of the particles we
use $r = (1+p)/4$ which is taken from minimum energy arguments
(e.g. Burbidge 1956\nocite{gb56}). Other typical observational
parameters are taken as follows. For the density profile of the
external atmosphere ($\rho _x = \rho _o (d/a_o) ^{-\beta}$) we take
the central density $\rho _o = 7.2 \cdot 10^{-22}$ kg/m$^3$, for a
core radius $a_o = 2$ kpc and $\beta = 1.9$; these values are typical
for a galaxy out to about 100 kpc from its center (Canizares {\em et
al.} 1987\nocite{cft87}). Rawlings \& Saunders (1991)\nocite{rs91}
find jet powers, $Q^{rs}$, from $10^{37}$ W to $10^{39}$ W for FRII
sources. These authors used in their calculations a form of the
minimum energy argument with cut--offs for the power law energy
distribution corresponding to the observational limits of the radio
spectrum of 10 MHz and 100 GHz respectively implying that there is no
thermal material present in the cocoon ($k=0$). Their frequency
cut--offs translate in our model to $\gamma _{i,min} \sim 500$ and
$\gamma _{i,max} \sim 10^5$ for a source with $D = 100 kpc$. Since in
our model the power law of the energy distribution extends to $\gamma
_{i,min} =1$ there is some thermal material present in the cocoon and
for the quoted values we find $k=3.6$. We must therefore adjust their
values to be consistent with our choice of $\gamma _{i,min}$ and
$\gamma _{i,max}$. Rawlings \& Saunders (1991)\nocite{rs91} also
assumed that half of the energy that is transported by the jet during
its life time, $t$, is lost to the surrounding IGM because of the
expansion work of the cocoon. However from KA97 we find that this work
is given by

\begin{equation} \int p_c \, dV_c = \frac{5-\beta}{9 \left[ \Gamma _c
+ (\Gamma _c -1) R^2 \right] 
-4 -\beta} \, Q_o \, t, \end{equation}

\noindent where $V_c$ is the volume of the cocoon. We can therefore
derive jet kinetic powers from the values given by Rawlings \&
Saunders (1991)\nocite{rs91} using

\begin{eqnarray} Q_o & = & \frac{9 \left[ \Gamma _c + ( \Gamma _c -1) R^2
\right] -4 - \beta}{10-2\beta} \,
(k+1)^{\frac{4}{5+p}} \, Q^{rs} \nonumber \\
& \approx & 13 \, Q^{rs},
\label{cowo}\end{eqnarray}

\noindent where $R=2$, $k=3.6$ and $\Gamma _c =5/3$. 

The pressure in the cocoon also depends on its axial ratio, $R$, and
the bulk velocity in the jet, $v_j$. We adopt $R=2$ which is an
average value (Leahy \& Williams 1984\nocite{lw84}) and $v_j =0.87c$
implying a Lorentz factor of the bulk motion of $\gamma _j =2$. This
velocity is a rather arbitrary choice since there are no reliable
estimates of $v_j$ in the literature.

The distribution of sources in the P--D diagram is constructed from
the data presented by Laing {\em et al.} (1983)\nocite{lrl83},
Subrahmanyan {\em et al.} (1996)\nocite{ssh96} and Cotter {\em et al.}
(1996)\nocite{crs96}; these authors quote luminosities at or close to
an observing frequency of 178 MHz and we therefore use this frequency
in our calculations. Note that the distribution of observed sources
presented here cannot be taken as the `true' distribution because of
the involved selection effects. The observed linear size of a radio
source projected onto the plane of the sky, $D$, depends of course not
only on the advance speed of the hot spots but also on $\sin \alpha$,
where $\alpha$ is the angle between the jet axis and the line of
sight. Assuming that the low--frequency LRL sample represents a
uniform distribution in $\alpha$ we adopt the average viewing angle
$\alpha = 39.5 ^{\circ}$ for the model sources.

\begin{figure}
\epsfig{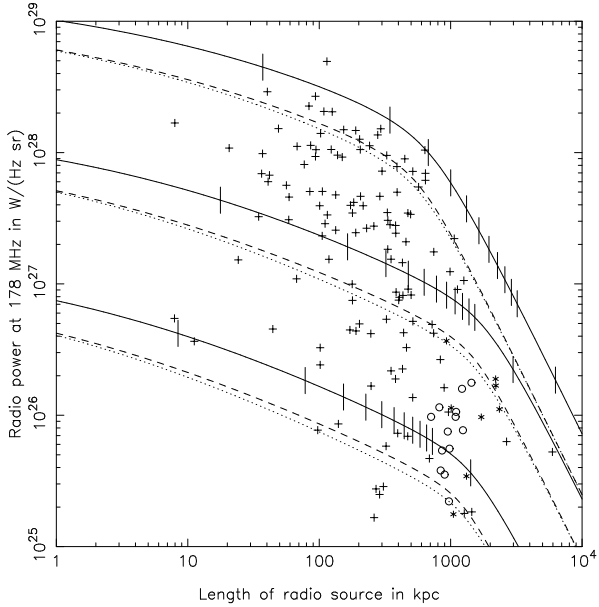}
\caption{Evolutionary tracks for three FRII
sources. The upper curves are for $Q_o = 1.3\cdot 10^{40}$ W and
$z=2$, the curves in the center for $Q_o =1.3\cdot 10^{39}$ W and
$z=0.5$ and the lower curves for $Q_o =1.3 \cdot 10^{38}$ W and
$z=0.2$. The other parameters for all sets of curves are: $R=2$,
$\rho_o = 7.2\cdot 10^{-22}$ kg/m$^3$, $a_o = 2$ kpc, $\beta =1.9$ and
$k'=0$. Solid curves for case 1, dashed curves for case 2 and dotted
curves for case 3 (see text). The vertical lines on the solid curves
are time markers. The first marker on the left indicates a life time
of $10^6$ years of the respective source model. Next ten markers for
every $10^7$ years up to $10^8$ years. At the last marker the source
is $2 \cdot 10^8$ years old. Crosses are observed sources above
10$^{25}$ W Hz$^{-1}$ sr$^{-1}$ taken from Laing et al. (1983), stars
from Subrahmanyan et al. (1996) and circles from Cotter et
al. (1996).}
\label{fig:modcom}
\end{figure}

\subsection{Evolution of the magnetic field\label{EVOL}}

To proceed we need to know the equation of state of the material in
the cocoon as a whole and also how the magnetic field is behaving
during the expansion of the volume elements $\delta V$. Since there is
a mixture of three different fluids in the cocoon this is not straight
forward, and it is also likely that the equation of state varies both
within the cocoon and with time. Hence we investigate three limiting
cases:

\begin{itemize}
\item Case 1: Both, the cocoon and the magnetic field energy density,
have a relativistic equation of state ($\Gamma _c = \Gamma
_{\scriptscriptstyle B} = 4/3$).
\item Case 2: The cocoon is `cold' ($\Gamma _c = 5/3$) but the energy
density of the magnetic field is proportional to the one of the
relativistic particles and therefore $\Gamma _{\scriptscriptstyle B} =
4/3$.
\item Case 3: Cocoon and magnetic field are `cold' ($\Gamma _c = \Gamma
_{\scriptscriptstyle B} = 5/3$). This implies some energy dissipation
process between the magnetic field and the non--relativistic
particles by which the adiabatic index of the energy density of the
magnetic field is held constant at 5/3. This effectively keeps the
energy of the magnetic field and that of the particles in
equipartition. 
\end{itemize}

In cases 1 and 3 the ratio of the energy densities of the magnetic
field and of the particles for every volume element $\delta V$ in the
cocoon, $r=0.785$ for the value of $p=2.14$ adopted here, is not
changing with time. In case 2 the value of $r$ will increase with time
since the adiabatic index of the magnetic field is lower than the one
of the particles. This implies that after some time the energy density
of the magnetic field will start to dominate the total energy density
in $\delta V$ and therefore change the adiabatic index of this part of
the cocoon. For a volume element injected into the cocoon at time
$t_i$ the ratio of energy densities $r$ becomes equal to 1 at time
$t\approx 1.9\cdot t_i$. At the same time the total volume of the
cocoon, $V_c$, has increased by a factor 6.4 ($V_c \propto
t^{9/(5-\beta)}$, see KA97) implying that in most parts of the cocoon
$r<1$. We therefore make the approximation that the adiabatic index of
the cocoon as a whole is $5/3$.

Figure \ref{fig:modcom}\nocite{lrl83}\nocite{ssh96}\nocite{crs96}
shows the evolutionary tracks for all three cases (corrected for
$\Gamma _c =5/3$: $Q_o = 13 \cdot Q^{rs}$) for two limiting jet
powers $Q_o = 1.3\cdot10^{38}$ W and $Q_o = 1.3\cdot10^{40}$ W and an
intermediate case with $Q_o = 1.3\cdot10^{39}$ W; the redshift $z$ is
0.2 for the low power jet, 0.5 for the `average' jet and 2 for the
high power jet.

The range of luminosities observed in FRII sources is bounded by the
tracks for the two limiting jet powers and the drop in luminosity
caused by the catastrophic energy losses of the relativistic electrons
occurs at approximately the correct linear size. The importance of
radiative losses in reducing the source luminosity at large linear
sizes (even at a low observing frequency) is very significant --- the
lack of large (Mpc) sources at high luminosity is therefore a direct
result of their intrinsic luminosity evolution, and is a strong
function of frequency. Most of the observed sources `crowd' in the
region of the diagram between 100 kpc and 1 Mpc. From the time markers
on the tracks in figure \ref{fig:modcom} it is clear that the model
discussed here predicts exactly this behaviour since model sources
spend the longest part of their life time in this region of the
diagram. The difference between cases 2 and 3 is almost
negligible. The higher luminosity for case 1 results from the fact
that in this case less of the energy transported by the jet is lost in
the expansion of the cocoon. Using the appropriate correction factor
for $Q_o$ for this case by setting $\Gamma _c=4/3$ in equation
(\ref{cowo}), which yields $Q_o = 2.3 \cdot Q^{rs}$, puts this track
very close to the other two. Thus we can conclude that the different
possibilities for the equation of state of the cocoon and for the
evolution of the magnetic field do not have a significant influence on
the evolutionary tracks. We will therefore use only case 3 in the
following.

\begin{figure}
\epsfig{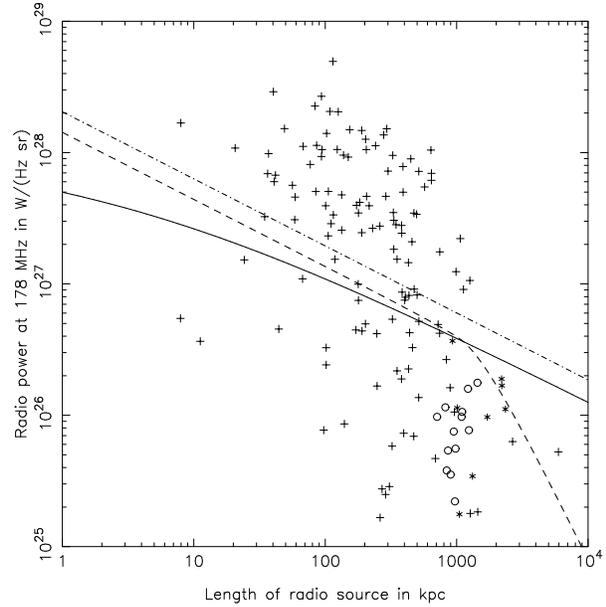}
\caption{Comparison of the different loss processes. All model
parameters as for the intermediate case in figure
\ref{fig:modcom}. The energy loss processes of the relativistic
electrons considered are different for the tracks: Solid curve:
adiabatic losses plus synchrotron losses, dashed: adiabatic losses
plus inverse Compton losses, dot--dashed: no losses.}
\label{fig:nonrad}
\end{figure} 

Figure \ref{fig:nonrad} shows a comparison of the effects of the
various loss processes. The most luminous track in this figure
represents a model where the energy losses of the relativistic
electrons are completely neglected, i.e. the right hand side of
equation (\ref{lora}) is set to zero. This curve is equivalent to the
results of Baldwin (1982)\nocite{jb82}. The two other tracks shown
both incorporate adiabatic losses plus one of the radiative loss
processes each. Obviously the adiabatic losses lead to a pure
luminosity off--set which one would expect from equation (\ref{lora})
if the two radiative terms on the right are set to zero. From the
figure it is also clear that synchrotron losses are dominating the
shape of the track for small sources and inverse Compton losses are
responsible for the steep decline of the luminosity of large
sources. The decreasing importance of synchrotron losses with
increasing linear size is caused by the decreasing energy density of
the magnetic field in the cocoon. Once the energy density of the CMBR
becomes comparable to the energy density of the magnetic field the
inverse Compton losses determine the shape of the track and limit the
size of the radio source.

As the source luminosity decreases as the result of inverse Compton
losses, so the prominence of the hot spots will increase. A detailed
investigation of this effect requires a proper model for the hot spot
itself and will be investigated in a forthcoming paper.

\subsection{Effects of redshift and environment}

\begin{figure}
\epsfig{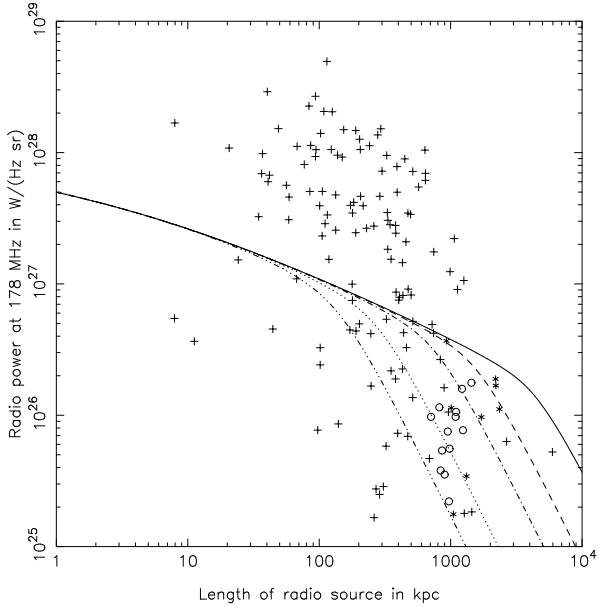}
\caption{The influence of the redshift $z$ on the evolutionary tracks
at $\nu =178$ MHz. Solid curve: $z=0$, dashed: $z=0.5$, dot--dashed:
$z=1$, dotted: $z=2$ and dot--dot--dot--dashed: z=3. All other model
parameters as in the intermediate case in figure \ref{fig:modcom}.}
\label{fig:redcom}
\end{figure}

In figure \ref{fig:redcom} we show evolutionary tracks at an observing
frequency of 178 MHz at various redshifts; the other source parameters
are given in the figure caption. These tracks are calculated assuming
that there is no evolution of the surrounding gas with redshift. Since the
energy density of the CMBR increases with redshift the inverse Compton
losses become important earlier and the evolutionary tracks begin to
steepen for smaller linear sizes of the source.

\begin{figure}
\epsfig{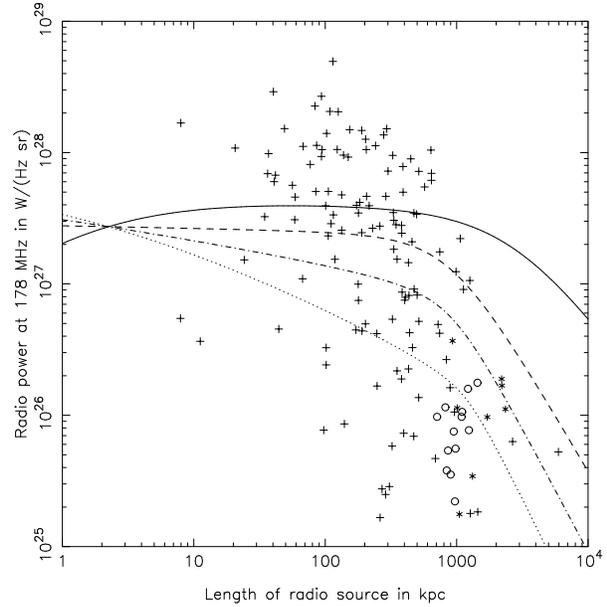}
\caption{The influence of the shape of the density distribution in the
IGM. Solid curve: $\beta =0$, dashed: $\beta =1$, dot--dashed: $\beta
=1.5$ and dotted: $\beta =2$. All other model parameters as in the
intermediate case in figure
\ref{fig:modcom}.}
\label{fig:betacom}
\end{figure}

The environments of radio sources have a great influence on their
appearance. The effects on the radio luminosity of a source caused by
a change in the shape of the density profile is shown in Figure
\ref{fig:betacom}, where the exponent, $\beta$, in the density
distribution of the IGM is varied. The track for $\beta = 0$ is purely
hypothetical since KA97\nocite{ka96b} have pointed out that a jet with
the parameters used here could not reach a linear size greater than
about 80 kpc in an uniform environment without being destroyed by
turbulence; for this value of $\beta$ the track is increasing in radio
luminosity for small sources, a result also found by Baldwin
(1982)\nocite{jb82}, but the inverse Compton losses are still strong
enough to reverse this trend and cause the track to steepen for large
linear sizes. The behaviour for $\beta =1$ is close to that of a
source with constant radio luminosity in the absence of radiative
losses.

The point of intersection is due to the intrinsic characteristic time
scale of the problem $\tau = (a_o^5 \, \rho _o / Q_o) ^{1/3}$ (see
KA97\nocite{ka96b}) which is the same for all values of $\beta$. Note
also that the point at which inverse Compton losses become important
shifts to smaller linear sizes for smaller values of $\beta$ (with the
exception of $\beta =0$). This is because the pressure in the cocoon is
lower at any given linear size in the case of large values of $\beta$
compared to smaller values. The population of relativistic electrons
in these cases has therefore lost less energy when the energy density
of the CMBR becomes comparable to the energy density of the magnetic
field in the cocoon. The steepening of the track due to inverse
Compton losses then occurs at larger linear sizes. In the case of
$\beta =0$ the luminosity of the source would increase with linear
size for all sizes without any radiative losses. However, the
synchrotron losses of the electrons in this case are so severe that
the luminosity of the source is almost entirely produced by material
which has just been injected into the cocoon while older parts are not
radiating anymore, leading to almost constant radio luminosity for all
linear sizes. Because of this the exact point at which inverse Compton
losses become important in this case is difficult to determine and the
start of the steepening of the track is less obvious.

The observed distribution of sources in the P--D diagram seems to be
fitted best by tracks with $\beta$ just less than 2. This is in
agreement with $\beta =1.9$ found by Cotter (1996)\nocite{gc96}, who
constructed evolutionary tracks through the P--D diagram by comparing
the positions of sources with similar observed jet power, and also with
X--ray observations of the gas surrounding galaxies at low redshift
(Canizares {\em et al.} 1987\nocite{cft87}).

\begin{figure}
\epsfig{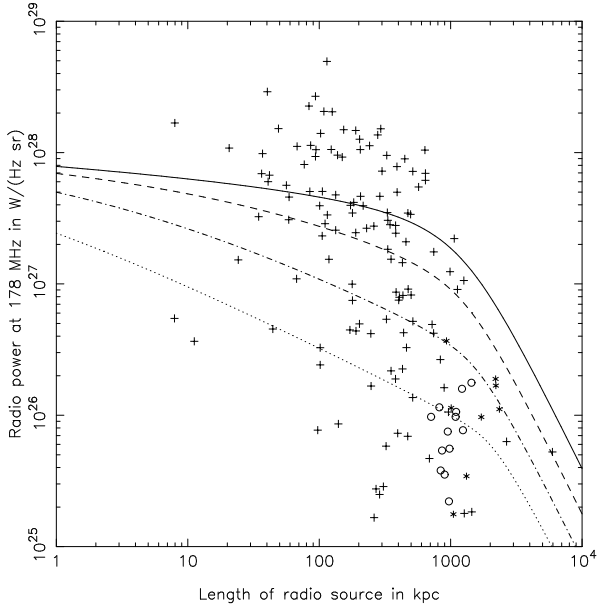}
\caption{The influence of the central density of the density
distribution of the IGM. Solid curve: $\rho_o = 7.2\cdot 10^{-20}$
kg/m$^3$, dashed: $\rho_o = 7.2\cdot 10^{-21}$ kg/m$^3$, dot--dashed:
$\rho_o = 7.2\cdot 10^{-22}$ kg/m$^3$ and dotted: $\rho_o = 7.2\cdot
10^{-23}$ kg/m$^3$. All other parameters as in the intermediate case
in figure \ref{fig:modcom}.}
\label{fig:rhocom}
\end{figure}

Figure \ref{fig:rhocom} shows the effect of varying the central density
of the IGM, $\rho _o$. For a greater central density the pressure in
the cocoon is higher which leads to an increased radio luminosity and
also to greater energy losses of the relativistic electrons via
synchrotron radiation. This effect `flattens' the tracks of small
sources in a high density environment. Again the steepening of the
tracks of large sources occurs at smaller linear sizes for sources in
denser environments since their hot spot advance speeds are smaller.

\subsection{Effects of jet properties}

\begin{figure}
\epsfig{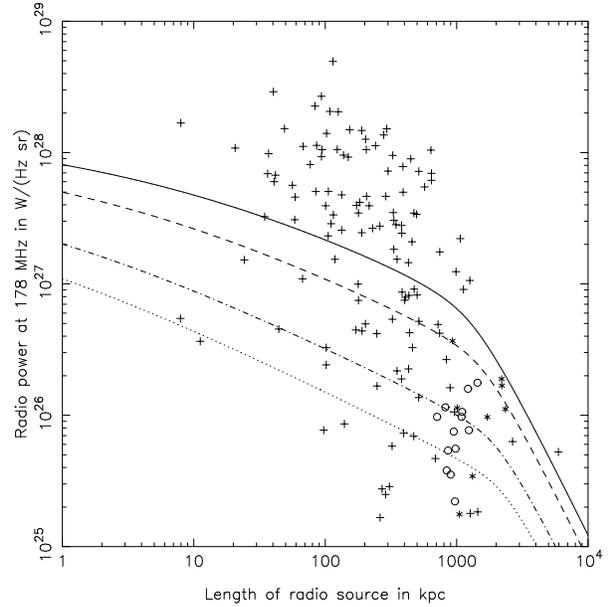}
\caption{The influence of the aspect ratio of the cocoon $R$. Tracks
are plotted for $Q_o = 1.3\cdot10^{39}$ W, $\rho_o = 7.2\cdot 10^{-22}$
kg/m$^3$, $a_o = 2$ kpc, $\beta = 1.9$, $k'=0$ and $z=0.5$. Solid
curve: $R = 1.3$, dashed: $R = 2$, dot--dashed: $R = 4$ and
dotted: $R = 6$.}
\label{fig:rtcom}
\end{figure}

In section \ref{EVOL} we have already investigated the influence of
the jet power, $Q_o$, on the evolutionary tracks. Other jet parameters
which have an effect on the radio luminosity of radio sources are the jet
half--opening angle, $\theta$, (which controls the aspect ratio of the
cocoon $R$) and the nature of the jet material. 

If the expansion of the cocoon perpendicular to the jet axis is
confined by the ram pressure of the IGM we find that the aspect ratio
of the cocoon is determined by the ratio of the pressure at the hot
spot to the pressure in the cocoon. From KA97\nocite{ka96b} we find
that this ratio is proportional to the inverse square of the jet half
opening angle $\theta$. Figure \ref{fig:rtcom} shows the evolutionary
tracks for the range of $R$ observed in sources at constant
$Q_o$. The limiting cases $R =1.3$ and $R=6$ correspond to
$\theta=47.8^{\circ}$ and $\theta=10.4^{\circ}$ respectively. The
pressure in the cocoon must be higher in the sources with wider
cocoons to allow the faster expansion perpendicular to the jet axis,
leading to higher synchrotron losses and hence giving relatively flat
tracks for small linear sizes. Since the energy density at the hot
spots is lower relative to that of the cocoons, the hot spot advance
speeds are smaller than those for the sources with thinner cocoons and
the steepening of the tracks occurs at smaller linear sizes.

\begin{figure}
\epsfig{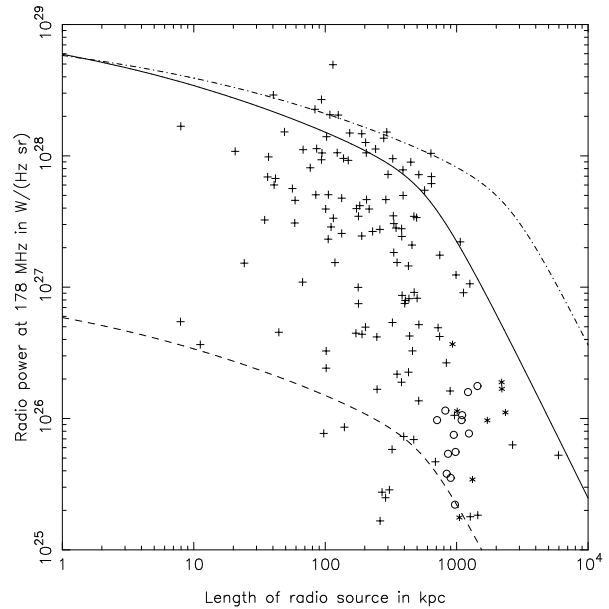}
\caption{The influence of protons in the jet. Tracks are plotted for
$\rho_o = 7.2\cdot 10^{-22}$ kg/m$^3$, $a_o = 2$ kpc, $\beta = 1.9$ ,
$z=2$ and $R = 2$. Solid curve: $Q_o = 1.3\cdot 10^{40}$ W and $k' =
0$, dashed: $Q_o = 1.3\cdot 10^{40}$ W and $k' = 100$ and dot--dashed:
$Q_o = 10^{42}$ W and $k' = 100$.}
\label{fig:kcom}
\end{figure}

All models considered up to this point have assumed that the jets
consist of an electron--positron plasma and that the contribution of
the thermal particles to the energy density in the cocoon is
negligible. If there are protons in the jet they are also accelerated
to relativistic velocities at the jet shock. Bell (1978)\nocite{ab78}
showed that for an proton--electron plasma accelerated in a shock
front the protons can store ten times more energy than the
electrons. Other acceleration scenarios predict much higher values
(Eilek \& Hughes 1991\nocite{eh91}). From figure \ref{fig:kcom} it is
clear that an addition of protons to the jet material, even if they
contribute only moderately to the energy density in the cocoon
($k'=100$), decreases the luminosity of radio sources severely. Should
the jets in FRII sources consist of protons and electrons they must
have much higher energy transport rates then we have assumed so far in
order to explain the most luminous sources. This would considerably
increase the hot spot advance speeds of these sources and lead in turn
to very large linear sizes for the sources before inverse Compton
losses become significant (see figure \ref{fig:kcom}). Since we do not
observe such large luminous sources this can be taken as evidence
against the presence of protons in the jet.

\section{Conclusions}

A model for the cocoon of FRII radio sources based on the model of
KA97 is presented. Energy loss processes for the relativistic
electrons producing the radio emission of the cocoon are incorporated
into a calculation of the expected radio luminosity. The resulting
evolutionary tracks through the P--D diagram are shown to be in good
agreement with the observed distribution of sources. The lack of
luminous giants in the diagram is reproduced just by the energy losses
of the electrons without invoking changes in the density distribution
of the IGM or the switching off of the central engine. The exact
details of the evolution of the magnetic field in the cocoon are shown
to have no significant effect on the evolutionary tracks. The effects
of the environment and jet parameters on the tracks are
investigated. The requirement of very high jet powers and the
associated high hot spot advance speeds for proton--electron jets to
account for the most luminous observed sources might rule out large
fractions of protons in the jet.\\[2.5ex]

\noindent {\bf ACKNOWLEDGEMENTS}\\[1ex]

We thank the referee Dr. J. P. Leahy for helpful comments on the
manuscript. 
 
\bibliography{/home/crk/tex/crk}
\bibliographystyle{../../mnras}

\end{document}